\documentclass[aps,prb,twocolumn,superscriptaddress,floatfix,nobibnotes]{revtex4-1}

\usepackage[utf8x]{inputenc}
\usepackage{amsmath,subfigure}
\usepackage[pdftex]{graphicx}
\usepackage{textcomp}
\usepackage[version=3]{mhchem} 

\usepackage{float} 
\floatstyle{boxed}

\usepackage{hyperref} 
\hypersetup{
    colorlinks=true,       
    linkcolor=red,          
    citecolor=blue,      
    filecolor=magenta,  
    urlcolor=blue,           
    urlbordercolor={0 1 1}}	

\newcommand{\degree}{$^{\circ}$}

\begin{document}

\newfloat{copyrightfloat}{thp}{lop}
\begin{copyrightfloat}
\raggedright
The peer reviewed version of the following article has been published in final form at  J. Phys. Chem. B, 2010, 114, 14426, doi: \href{http://dx.doi.org/10.1021/jp1012347}{10.1021/jp1012347}.
\end{copyrightfloat}


\title{Spin Signatures of Photogenerated Radical Anions in Polymer--[70]Fullerene Bulk Heterojunctions: High Frequency Pulsed EPR Spectroscopy}

\author{Oleg G. \surname{Poluektov}}\email{Oleg@anl.gov}
\affiliation{Chemical Sciences and Engineering Division, Argonne National Laboratory, 9700 Cass Avenue, Argonne, Illinois 60439}
\author{Salvatore \surname{Filippone}}
\author{Nazario \surname{Mart{\'i}n}}
\affiliation{Departamento de Qu{\'i}mica Org{\'a}nica, Facultad de Ciencias Qu{\'i}micas, Universidad Complutense de Madrid, 28040, Madrid, Spain, IMDEA-Nanociencia, Campus Universidad Aut{\'o}noma de Madrid, Spain}
\author{Andreas \surname{Sperlich}}
\author{Carsten \surname{Deibel}}
\author{Vladimir \surname{Dyakonov}}\email{Dyakonov@physik.uni-wuerzburg.de}
\affiliation{Julius-Maximilians University of W\"urzburg and Bavarian Centre for Applied Energy Research e. V. (ZAE Bayern), D-97074 W\"urzburg, Germany}


\begin{abstract}
Charged polarons in thin films of polymer--fullerene composites are investigated by light-induced electron paramagnetic resonance (EPR) at 9.5 GHz (X-band) and 130 GHz (D-band). The materials studied were poly(3-hexylthiophene) (PHT), [6,6]-phenyl-C61-butyric acid methyl ester (\ce{C60}-PCBM), and two different soluble \ce{C70}-derivates: \ce{C70}-PCBM and diphenylmethano[70]fullerene oligoether (\ce{C70}-DPM-OE). The first experimental identification of the negative polaron localized on the \ce{C70}-cage in polymer--fullerene bulk heterojunctions has been obtained. When recorded at conventional X-band EPR, this signal is overlapping with the signal of the positive polaron, which does not allow for its direct experimental identification. Owing to the superior spectral resolution of the high frequency D-band EPR, we were able to separate light-induced signals from P$^{+}$ and P$^{-}$ in PHT--\ce{C70} bulk heterojunctions. Comparing signals from \ce{C70}-derivatives with different side-chains, we have obtained experimental proof that the polaron is localized on the cage of the \ce{C70} molecule.
\end{abstract}


\maketitle

\section{Introduction}
Photovoltaic (PV) cells are the most promising man-made devices for direct solar energy utilization. In analogy with natural photosynthesis, the key steps in PV solar energy conversion are the generation, separation, and extraction of charges. PV systems can be classified into three big groups based on the active media: inorganic, organic, and hybrid devices. In spite of the fact that organic based PVs did not demonstrate competitively high conversion efficiency so far, they are considered as a high potential option with many attractive features like low cost fabrication and tunability of electronic properties of organic materials. Recently, a considerable improvement of the efficiency in the range of 8\% has been recorded for PVs based on conjugated polymer--fullerene composites~\cite{Green:2010km}. In general, by blending the conjugated polymer and fullerenes, efficient light-induced charge separation (CS) can be achieved, thus making these materials attractive for solar energy conversion~\cite{Sariciftci:1992wb,MORITA:1992th}.

As it was demonstrated for natural photosynthesis, an extremely effective technique for investigation of the light-induced generation, separation, and recombination of the charge carriers is the advanced electron paramagnetic resonance (EPR) technique, such as light-induced, multifrequency, time-resolved, pulsed, high frequency EPR~\cite{Borbat:2001fq,Poluektov:2005jc}. Light-induced EPR (LEPR) has been extensively used to characterize the electronic structure of the charges in composites of poly[2-methoxy-5-(3',7'-dimethyloctyloxy)-1,4-phenylenevinylene] and [6,6]-phenyl-\ce{C61}-butyric acid methyl ester (\ce{C60}-PCBM), a soluble derivative of \ce{C60}~\cite{Dyakonov:1999ub}. Under illumination of the sample, two paramagnetic species are formed due to photoinduced CS between conjugated polymer and fullerene. They are the positive, P$^{+}$, and the negative, P$^{-}$, polarons on the polymer backbone and \ce{C60}-cage, respectively. EPR spectra of these species were completely resolved and characterized using high spectral resolution of the high frequency (HF) EPR \cite{Ceuster:2001hc,Aguirre:2008bw}.

One of the downsides of the \ce{C60}-PCBM acceptor material for photovoltaic applications is a very low absorption coefficient in the visible spectral region and, as a consequence, a relatively small contribution to the photocurrent. The substitution of the \ce{C60} by \ce{C70} fullerene considerably improves the photocurrent due to the low symmetry of the \ce{C70} molecules and thus higher absorption coefficient in the visible region~\cite{Wienk:2003kl}.

While the understanding of the elementary steps of the efficient charge separation and charge stabilization in the photovoltaic materials is a prerequisite for improving the efficiency of the organic PV cells, there is little known on the photophysics in \ce{C70} containing composites. VIS-NIR spectra of \ce{C70} anions obtained by reduction using potassium in tetrahydrofuran solution as well low frequency (LF) EPR studies were reported more than a decade ago~\cite{Gherghel:1995um}. Recent studies of fullerene \ce{C70} and \ce{C60} dimers revealed additional photoinduced absorption and LF EPR features due to \ce{C70} fullerenes~\cite{Delgado:2009bt}.

Here, we report on the photoinduced CS state in polymer--\ce{C70}-fullerene composites. This is the first characterization of the CS polaron states (radical anion and cation) in \ce{C70}-based polymer--fullerene heterojunctions by high frequency (130 GHz) LEPR. Having high spectral resolution, the HF LEPR allows identification of the spectra of the positive polaron localized on the polymer and the negative one on the fullerene cage.

The LEPR spectra of the polymer--\ce{C70}-fullerene blends studied with the conventional X-band (9.5 GHz) technique were found to be significantly different from those of polymer--\ce{C60}-PCBM and poorly resolved, which makes the direct identification of the CS species speculative. To clarify this issue, we have carried out a comparative high and low frequency LEPR study of polymer--fullerene composites.

\section{Experimental Section}
\begin{figure}[ht]
 \centering
	\includegraphics[width=\columnwidth]{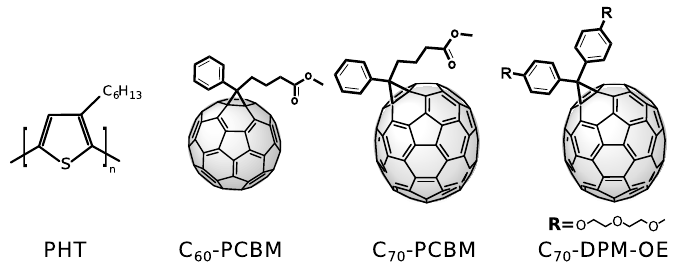}
	\caption{Structures of PHT and of \ce{C60}- and \ce{C70}-Derivatives}
	\label{fig:mat}
\end{figure}

The polymer used in this work was the regioregular poly(3-hexylthiophene) (PHT, Rieke Metals), and the fullerenes were soluble derivatives \ce{C60}- and \ce{C70}-PCBM (Solenne B.V.). We have also synthesized and studied a new compound \ce{C70}-DPM-OE, in which each phenyl group of the diphenylmethanofullerene (DPM) moiety is decorated with a solubilizing oligoether (OE) chain. The synthesis has been carried out as reported for related compounds~\cite{Riedel:2005ec} (see the Supporting Information). The chemical structures of the materials studied are shown in Figure \ref{fig:mat}.

The PHT--fullerene composites were prepared from toluene and chlorobenzene solutions of the PHT--fullerene mixture (1:1 weight ratio) by (1) drying sample in the N$_2$ atmosphere, (2) removing the rest of the solvent by pumping samples for 48~h, (3) annealing the sample in N$_2$ atmosphere at 130~\degree C for 10~min. The LEPR experiments were carried out with continuous wave (cw) X-band (9.5~GHz/0.33~T) EPR (modified Bruker 200D) by using excitation of 532~nm with a cw diode-pumped solid state laser. By using the static magnetic field modulation and the lock-in techniques, cw EPR spectra were recorded as the first derivative of the absorption. High frequency EPR measurements were performed on a pulsed high frequency D-band (130~GHz/4.6~T) spectrometer equipped with a single mode cylindrical cavity TE$_{011}$~\cite{Poluektov:2002be}. EPR spectra of the composites were recorded in the pulse mode in order to get rid of the microwave phase distortion due to fast-passage effects. The absorption line shape of the EPR spectra was recorded by monitoring the electron spin echo (ESE) intensity from a two microwave pulse sequence as a function of magnetic field. The duration of the $\pi$/2 microwave pulse was 40--60~ns, and typical separation times between microwave pulses were 150--300~ns. The differential line shapes of the spectra were obtained by numerical differentiation. Light excitation of the sample was achieved with an optical parametric oscillator (Opotek) pumped by a Nd:YAG laser (Quantel), the output of which was coupled to an optical fiber. The optical fiber allows delivery of up to 2~mJ per pulse to the sample. The excitation wavelength was 532~nm. The g-tensor parameters were obtained from theoretical simulation of the EPR spectra using SIMFONIA and EasySpin software~\cite{Stoll:2006ks}.

\section{Results and discussion}

\begin{figure}[ht]
 \centering
	\includegraphics[width=.7\columnwidth]{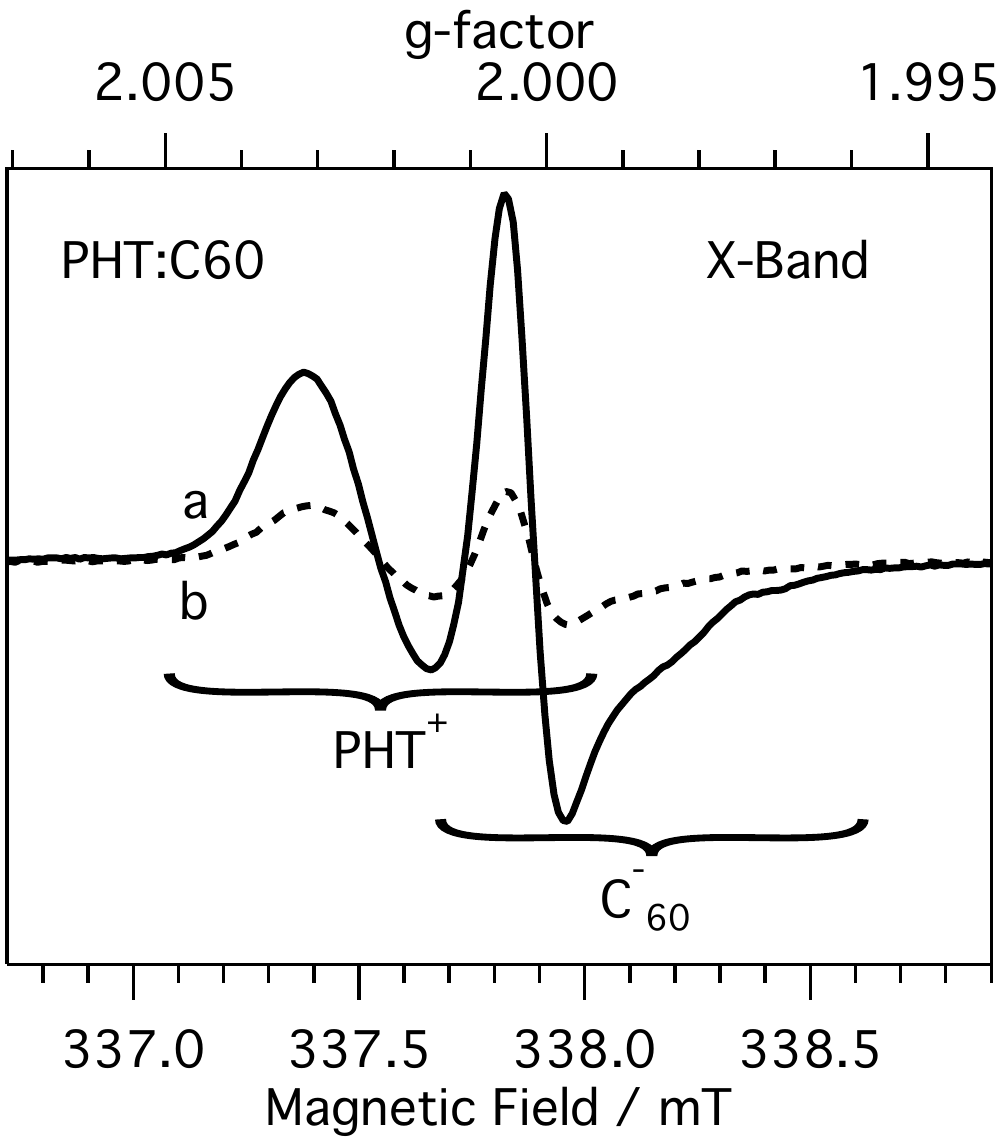}
	\caption{X-band (9.5 GHz/0.33 T) LEPR spectra of PHT--\ce{C60} composites, $\lambda_{exc}$ = 532~nm, T = 30~K: (a) under illumination; (b) after illumination (light off). Note: spectra are recorded as the first derivative of the microwave absorption.}
	\label{fig:C60-XESR}
\end{figure}

\begin{figure}[ht]
 \centering
	\includegraphics[width=\columnwidth]{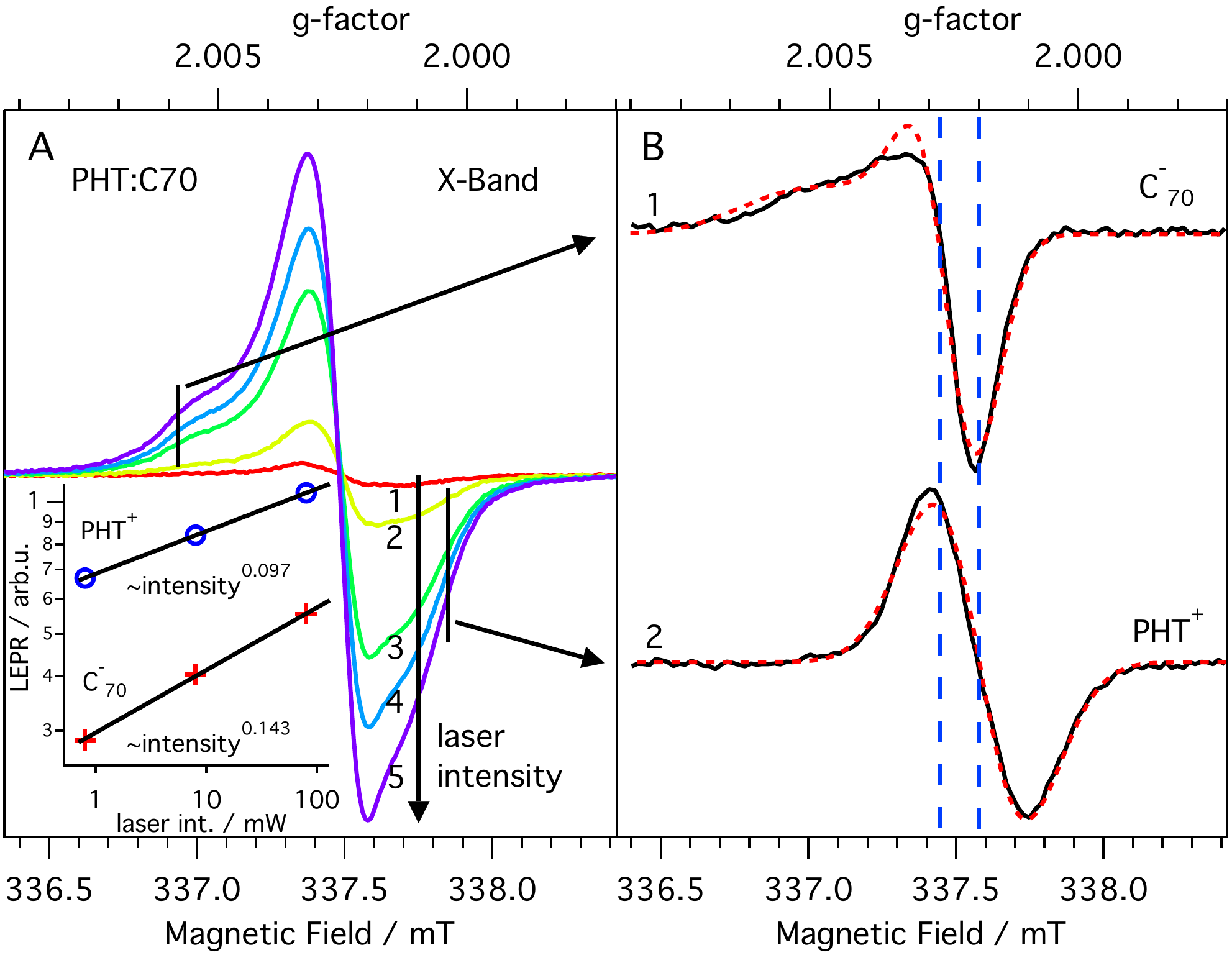}
	\caption{X-band (9.5~GHz / 0.33~T)LEPR spectra of PHT / \ce{C70}-PCBM composites at  T = 30~K. (A) Experimental spectra 1--5 taken under different illumination conditions: dark, ambient lab illumination, 0.8, 8, and 80 mW laser intensity. $\lambda_{exc}$ = 532~nm. The vertical markers indicate the positions (shoulders) at which the light intensity dependence was measured. Inset: LEPR amplitude as a function of laser intensity for the marked position corresponding to \ce{C70}$^{-}$ and PHT$^{+}$ contributions to the spectra and fitted with a power law. (B) Subtracted spectra of A2 and A5 after normalization at the vertical markers are tentatively assigned to 1 (\ce{C70}$^{-}$) and 2 (PHT$^{+}$). Red (dashed) lines: simulated spectra with parameters from Table \ref{tab:5-C70-g-components}.}
	\label{fig:C70-XESR}
\end{figure}

\begin{figure}[ht]
 \centering
	\includegraphics[width=.7\columnwidth]{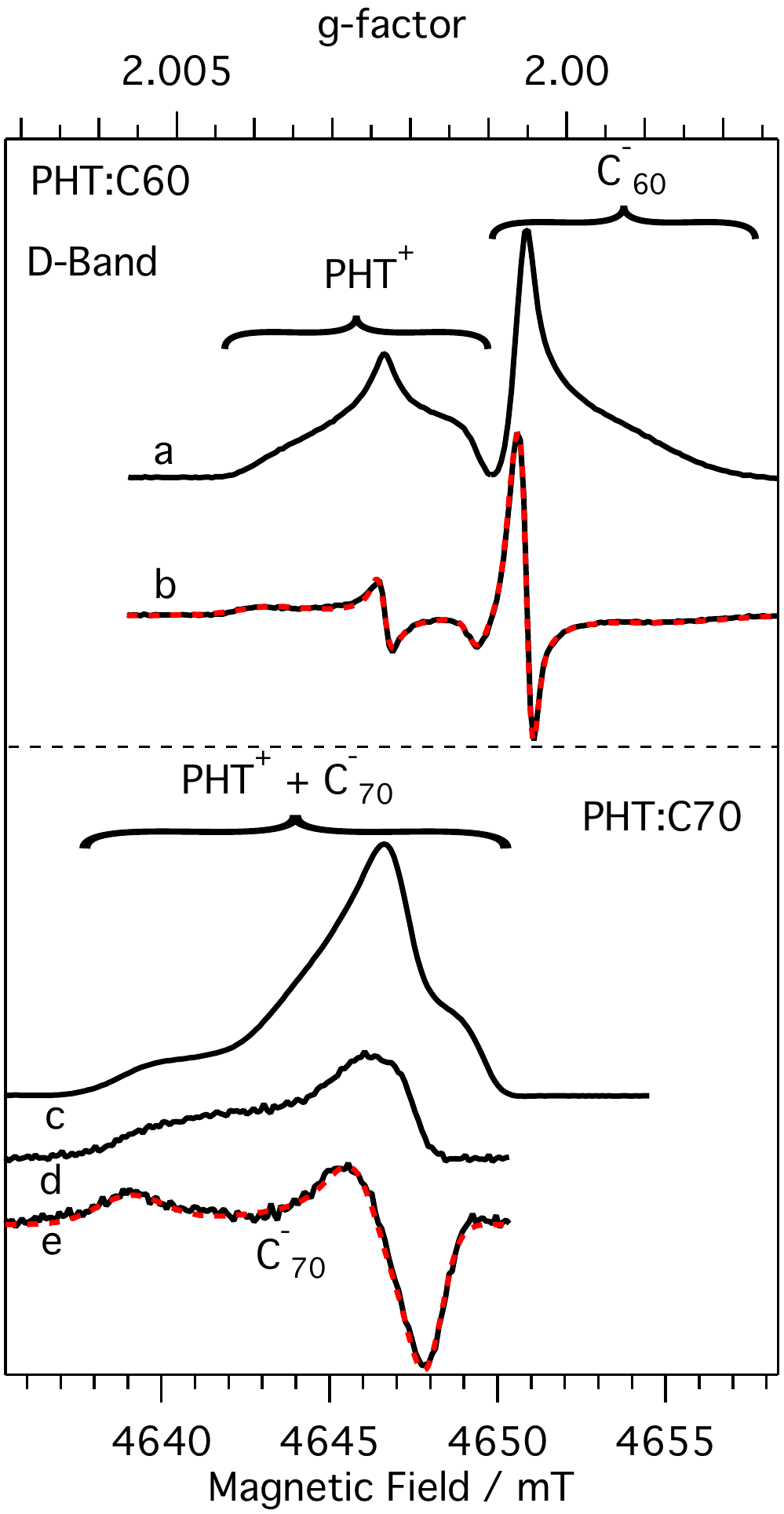}
	\caption{HF (130~GHz/4.6~T) EPR spectra of PHT--\ce{C60} (a, b) and PHT--\ce{C70} (c--e) composites. $\lambda_{exc}$ = 532~nm, T = 30~K. (a, c) Under illumination; (b) first derivative of a, red dashed curves - theoretical simulation; (d) subtraction of P$^+_{PHT}$ spectrum (low field part of a) from the spectrum of PHT--\ce{C70} composites (c); (e) first derivative of d, red dashed curves - theoretical simulation. The simulation parameters are in Table \ref{tab:5-C70-g-components}.}
	\label{fig:DESR}
\end{figure}

\begin{table}[h]
\caption{Main Components of the g-Tensors of Positive, P$^+$, and Negative, P$^-$, Polarons in PHT--Fullerene Blends}
\begin{center}
\begin{tabular}{c|ccc}
$g$-tensor$^{ ~a)}$ & P$^{+}$ in PHT   & P$^{-}$ in \ce{C60}-deriv. & P$^{-}$ in \ce{C70}-deriv. \\ 
\hline
$g_{xx}$   &  2.00380  & 2.00058     & 2.00592   \\
$g_{yy}$   &  2.00230  & 2.00045     & 2.00277   \\
$g_{zz}$   &  2.00110  & 1.99845     & 2.00211   \\
\hline
\end{tabular}
$^{a)}$ The relative error in the g-tensor measurements is $\pm$0.00005.
\end{center}
\label{tab:5-C70-g-components}
\end{table}%

Prior to illumination, no EPR signals in the blends were observed. Under illumination at temperatures below 150~K, the X-band spectra of the composites containing \ce{C60}-derivatives show two distinct polaron signals (Figure \ref{fig:C60-XESR}a). These signals were previously assigned to photogenerated positive and negative polarons in the blends: low field signal to P$^+$ localized on the polymer and high field signal to P$^-$ on the \ce{C60}-cage~\cite{Sariciftci:1992wb,MORITA:1992th,Dyakonov:1999ub,Ceuster:2001hc}. At lower temperatures, however, part of the signal remains persistent after switching off the excitation, as shown in Figure \ref{fig:C60-XESR}b and can be eliminated only by warming up the sample\cite{Dyakonov:1999ub}.

Remarkably, X-band spectra of \ce{C70}-based blends do not demonstrate any signal in the high field region (Figure \ref{fig:C70-XESR}). Instead, only one signal, corresponding to the low field signal in Figure \ref{fig:C60-XESR} and being attributed to P$^+$ (radical cation) on a PHT chain can be seen. It is slightly broadened and reveals a low field (B = 336.9~mT) shoulder. We assume the appearance of this shoulder is related to the negative polaron P$^-$ localized on \ce{C70}-PCBM.

In order to decompose strongly overlapping signals, the ``light on--light off'' method with subsequent annealing can be applied. In general, the LESR line intensity is proportional to the number of spins in the sample, which is in turn proportional to the illumination intensity. Figure \ref{fig:C70-XESR}A shows the spectral series taken in the dark, under ambient light illumination, at 0.8, 8, and 80 mW laser excitation powers, respectively. The vertical markers in Figure \ref{fig:C70-XESR}A indicate the positions (shoulders) of interest, at which the amplitudes of the spectra were determined. These values are shown in the inset to Figure \ref{fig:C70-XESR}A, indicating slightly different behavior for different parts of the EPR spectra. In particular, it is seen that, at the lowest excitation intensities, the low field shoulder is strongly suppressed. This finding was used for tentative decomposition of the LEPR to positive (PHT$^+$) and negative (\ce{C70}$^-$) polarons by normalization and subsequent subtraction, as shown in Figure \ref{fig:C70-XESR}B. The line shape and the g-value of the \ce{C70}$^-$ EPR spectrum obtained as described above considerably differ from those of \ce{C60}$^-$ in a similar composite environment. This striking observation is ambiguous, as the applied decomposition method is based on the assumption that \ce{C70}$^-$ and PHT$^+$ EPR signals have different light saturation behavior. Therefore, a direct experimental confirmation is necessary.

To separate overlapping signals with close g-values and to extract the parameters of the g-tensor, we applied the high field EPR technique. HF LEPR spectra of PHT--\ce{C60} and PHT--\ce{C70} are shown in Figure \ref{fig:DESR}. The P$^+$ and P$^-$ signals of PHT--\ce{C60}-PCBM composites are now completely separated (Figure \ref{fig:DESR}a). Note, in a pulsed EPR experiment, the signals are detected as microwave absorption, while, in cw EPR experiments, the signals are detected as the first derivative of the absorption. For better comparison, we depicted the first derivative EPR spectrum together with the fit in Figure \ref{fig:DESR}b. The signal at high field corresponds to P$^-$ and shows near axial symmetry of the g-tensor. The high field part of this signal (corresponding parallel component of the g-tensor of the P$^-$ spectrum) is very broad. Such large broadening is typical for g-strain effect, when one or another g-tensor component is distributed mainly owing to the interaction with the environment. In the first approximation, the bigger initial shift of the g-tensor component from the free electron g-tensor, g$_e$ = 2.0023, the stronger g-strain effect. The large g-strain of the g$_{par}$ component reflects the sensitivity of the \ce{C60}$^-$ molecular orbital energy levels of unoccupied molecular orbitals (see below) to the heterogeneous environment.

The low field line (Figure \ref{fig:DESR}a,b) is due to a positive polaron on the polymer chain with rhombic symmetry of the g-tensor. The low field component of P$^+$ is also unusually broadened which might be explained by the g-strain effect as well, and sensitivity of the PHT$^+$ energy levels of occupied orbitals (see below) to the heterogeneous environment.

The HF LEPR spectra of PHT--\ce{C70}-derivatives show only a broad signal at low field (Figure \ref{fig:DESR}c). This signal is much more extended toward the lower field than the signal of P$^+$ in PHT--\ce{C60}-derivative blends. Presumably, it consists of two overlapping spectra, one corresponding to P$^+$ and another one to P$^-$. As the line shape of the P$^+$ signal, which is localized on PHT, is expected to be identical for the \ce{C60} and \ce{C70} composites, we can obtain a spectrum of the radical anion P$^-$ localized on \ce{C70} by subtracting the low field parts of these two spectra, as shown in Figure \ref{fig:DESR}d. Again, to draw parallels between cw- and pulsed-EPR, the spectrum Figure \ref{fig:DESR}e is plotted as the first derivative of Figure \ref{fig:DESR}d together with the fit. The residual signal demonstrates the g-tensor close to axial symmetry, with higher anisotropy compared to the polaron localized on \ce{C60}, which reflects lower symmetry of the \ce{C70} molecule.

The signal obtained by the subtraction procedure is definitely due to a negative polaron localized on the \ce{C70}-PCBM molecule. However, it can be localized either on the \ce{C70}-cage or on the side-chain. To resolve this question, we recorded spectra from the composites of PHT--\ce{C70}-derivatives that differ by the side-chains (Figure \ref{fig:mat}), e.g., the \ce{C70}-DPM-OE. Both X- and D-band EPR spectra were found to be identical, i.e., did not depend on the chemical structure of the side-chain. Thus, we believe that the spectra in Figure \ref{fig:DESR} are due to negative polarons being localized on the \ce{C70}-cage. The previously identified by the subtraction technique charge transfer polaron P$^-$ (Figure \ref{fig:C70-XESR}B) is in good agreement with the spectrum measured directly by HF LESR. The g-tensor parameters of the polaronic signals are summarized in Table \ref{tab:5-C70-g-components}.

It is important to note that, while the components of the g-tensor of the \ce{C60} anion radical are lower than the free electron g-factor, g$_e$ = 2.0023, for \ce{C70}$^-$, the g-tensor components are mostly higher than g$_e$ = 2.0023. According to the classical Stone theory of g-factors~\cite{Stone:1963bk}, negative deviation of the g-factor from the free electron value is due to spin--orbit coupling with empty p- or d-orbitals, while spin--orbit coupling with occupied orbitals leads to positive deviations. The latter case is more often encountered for pure organic radicals. Thus, a difference in the g-value is an indication of the different electronic structure of molecular orbitals in \ce{C60} vs \ce{C70} anion radicals. Previous studies of \ce{C70}$^-$ in the liquid phase at room temperatures~\cite{Gherghel:1995um,Friedrich:1994fm} demonstrate that the average g-factor of \ce{C70}$^-$ is higher than the g-factor of \ce{C60}$^-$.This positive shift of the g-factors in solution for \ce{C70}$^-$ relative to \ce{C60}$^-$ has been explained on the basis of the static Jahn--Teller effect~\cite{Dubois:1991cu,Adrian:1996wm}. It is probable that Jahn--Teller dynamics in the solid phase is quite different for \ce{C60} and \ce{C70} molecules, which might contribute to different signs of the g-value shifts~\cite{Adrian:1996wm}. Nevertheless, there is no unified theory that can explain g-tensors of both \ce{C60}$^-$ and \ce{C70}$^-$ radicals yet. Our precise measurements of the anisotropic g-tensor of \ce{C70} anion radical in the solid phase might be a reference point for developing and testing such a unified theory.

Finally, the reported pulsed HF EPR experiments were carried out on samples that were dried in the N$_2$ atmosphere. Upon removing the solvent by evaporation, the intensities of the lines considerably decrease and almost completely disappear after annealing of the films. At the same time, narrowing of the P$^+$ spectra was observed for all samples. We believe that this effect is connected to phase separation of the polymers and fullerenes during which leads to the higher delocalization of the polarons. A high degree of the polaron delocalization in the annealed samples is also confirmed by shortening of the relaxation time T$_1$ as measured by the spin echo technique at HF EPR.

\section{Conclusion}
To summarize, we have provided the first experimental EPR identification of negative polaron (radical anion) localized on the \ce{C70}-cage in PHT--\ce{C70}-derivative composites using the LEPR technique. When recorded with conventional X-band LEPR, this signal is overlapping with the signal of the positive polaron, which does not allow for its direct identification. Owing to the superior spectral resolution of the HF (130 GHz) EPR technique, we were able to separate light-induced signals from P$^+$ and P$^-$. Comparing signals from \ce{C70}-derivatives with different side-chains, we got a confirmation that the polaron is localized on the cage of the \ce{C70} molecule. The obtained g-tensor parameters are of importance, as these are the characteristics of the structure, symmetry, and dynamics of the localized/delocalized unpaired spin states.

\begin{acknowledgments}
The work at ANL was supported as part of the ANSER, an Energy Frontier Research Center funded by the U.S. Department of Energy, Office of Science, Office of Basic Energy Sciences. The work at the University of W\"urzburg was supported by the German Research Foundation, DFG, within the SPP ``Elementary processes in organic photovoltaics'', under contract DY18/6-1. The MICINN of Spain (project CT2008-00795/BQU, R\&C program, and Consolider-Ingenio 2010C-07-25200) and the CAM (project P-PPQ-000225-0505) are also acknowledged. V.D. acknowledges financial support from ANSER during his research visit at ANL.
\end{acknowledgments}

\section*{Supporting Information Available}
Chemical synthesis and spectroscopic analysis of compounds. This material is available free of charge via the Internet at \href{http://pubs.acs.org/doi/suppl/10.1021/jp1012347}{http://pubs.acs.org}.

\bibliography{hex}

\end{document}